\documentclass[showpacs,preprintnumbers,amsmath,amssymb]{revtex4}
\usepackage{graphicx}
\usepackage{dcolumn}
\usepackage{bm}
\begin{document}
\title{\bf A New Type of Superstring in four dimensions}
\medskip
\author{B. B. Deo}
\affiliation{ Physics Department, Utkal University, Bhubaneswar-751004, India.}
\begin{abstract}
A bosonic string in twenty six dimensions is effectively 
reduced to four dimensions by eleven Majorana fermions which are vectors 
in the bosonic represetation SO(d-1,1). By dividing 
the fermions in two groups,  actions can be written down which are world sheet
supersymmetric, 2-d local  
and local 4-d supersymmetric. The novel string is anomally free, 
free of ghosts and the partition function is modular invariant.
\end{abstract}
\maketitle
There are many ways to construct four dimensional superstrings. Antoniadis, Bachas and
Kounnas~\cite{Int} have used trilinear coupling whereas Chang and Kumar~\cite{ch}
have taken Thirring coupling of Fermions in addition to four bosonic coordinates.
Kawai, Lewellen and Tye \cite{ka} have discussed the problem in a fairly general way. 
Consistent superstring as solutions of D=26 bosonic string exits as has been shown 
by Casher, Englert, Nicolai and Taormina \cite{cas} and this reasoning has been 
extended by Englert, Houart and Taormina \cite{en} to  brane fusion of the bosonic 
string to form a fermionic string. We introduce fermions by noting that one boson is 
equivalent to two fermions. In a formal way, one can add 44 world sheet scalar fermions 
with internal symmetry
SO(44) to the four bosonic 
coordinates to obtain an equivalent Nambu Goto string in four dimensions. In 
this letter we propose to effectively compactify the 26d string by 11 
worldsheet vector fermions. There will be eleven metric ghosts which can be 
eliminated by eleven subsidiary conditions. 
The action in the world sheet ($\sigma,\tau$) is
\begin{equation}
S= -\frac{1}{2\pi}\int d^2\sigma[\partial^{\alpha}X^{\mu}\partial_{\alpha}X_{\mu}
-i \sum^{11}_{j=1}\bar\phi^{\mu,j}\rho^{\alpha}\partial_{\alpha}\phi_{\mu,j}]\label{q}\\
\end{equation}
The internal symmetry of the  $\phi^{\mu}$'s is SO(11). They are real Majorana fermions
which are vectors in the bosonic representation SO(3,1). If the action is to be supersymmetric,
we must find a supersymmetric partner $\Psi^{\mu}$ to the bosonic coordinate $X^{\mu}$.
This can only be the linear sum of $\phi^{\mu}$'s like $\Psi^{\mu}=\sum_{j=1}^{11}
e^j_{\phi}\phi_j^{\mu}$. With usual anticommutator following from action (\ref{q}), the
$\Psi^{\mu}$ will have the same anticommutator rule provided $\sum_{j=1}^{11}e_{\phi}^j
e_{\phi j}=1$. We searched for such coefficients to find a linear combination which will 
make theaction supersymmetric. After extensive and laborious probing we found only one 
possibility namely to group $\phi^{\mu}$'s into two
species $\psi^{\mu,j}$: j=1,2....6 and $\phi^{\mu,k}$; k=1,2,..5. For one group, 
the positive and negative frequency are $\psi^{\mu,j }=\psi^{(+)\mu,j} + 
\psi^{(-)\mu,j}$ where
as for the other, allowing for the phase uncertainty of the creation operator
for Majorana spinors,  $\phi^{\mu,k} = \phi^{(+)\mu,k} - \phi^{(-)\mu,k}$.
For the latter species, the equal time anticommutator has a negative sign. 
The action is taken as
\begin{equation}
S= -\frac{1}{2\pi}\int d^2\sigma [ \partial_{\alpha}X^{\mu}\partial^{\alpha}X_{\mu}
-i \bar{\psi}^{\mu,j}\rho^{\alpha}\partial_{\alpha}  \psi_{\mu,j}
+ i \bar{\phi}^{\mu,k}\rho^{\alpha}\partial_{\alpha}  \phi_{\mu,k}]\label{s}\\
\end{equation}
The 2d Dirac matrices are
\begin{eqnarray}
\rho^0 =
\left ( 
\begin{array}{cc}
0 & -i\\
i & 0\\
\end{array}
\right )
\end{eqnarray}
and
\begin{eqnarray}
\rho^1 =
\left ( 
\begin{array}{cc}
0 & i\\
i & 0\\
\end{array}
\right )
\end{eqnarray}
and $\bar{\psi}=\psi^+\rho^o$, $\bar{\phi}=\phi^+\rho^o$\\
It is worth while noting that the light cone fermionic action
\[ S_{l.c.}=\frac{i}{2\pi}\;\int d^2\sigma \sum_{\mu =0,3}\left( \bar{\psi}^{\mu j}\rho^
{\alpha}\partial_{\alpha}\psi_{\mu j} - \bar{\phi}^{\mu k}\rho^
{\alpha}\partial_{\alpha}\phi_{\mu k}\right) \] has central charge 11 and $S - S_{lc}$ has 
central charge 15 as the usual N=1, supersymmetric action. In this model,$S_{l.c.}$ is the
equivalent of the Faddeev- Popov super conformal ghost action of the superstring.
It is necessary to introduce two unit vectors ${\bf{e}}_{\psi}$  and 
${ \bf{e}}_{\phi}$ with
eleven components e.g. for j=3, and k=3,
\begin{equation} 
{\bf{e}}_{\psi}^3=(0,0,1,0,0,0;0,0,0,0,0),\;\;\;\;\;  {\bf{e}}_{\phi}^3 =
(0,0,0,0,0,0;0,0,1,0,0)
\end{equation}
with properties ${e}^j_{\psi}{e}_{\psi {n}}=\delta^j_{n}$ and 
${e}^k_{\phi}{e}_{\phi m}=\delta^k_m$. We shall frequently use
 ${e}^j_{\psi}{e}_{\psi j}$=6 and
${e}^k_{\phi}{e}_{\phi k}$=5. Every where the upper index referes to a column and the lower
to a row. The suffixes $\psi$ and $\phi$ are for book keeping only.
The action S in (\ref{s}) is found to be invariant under the following 
infinitesimal supersymmetric transformation.
\begin{eqnarray}
\delta X^{\mu} =\bar{\epsilon}(e^j_{\psi}\psi^{\mu}_j - e^k_{\phi}\phi^{\mu}_k),
\nonumber\\
\delta\psi^{\mu,j}= - ie^j_{\psi}\rho^{\alpha}\partial_{\alpha}X^{\mu}\epsilon\nonumber\\
\end{eqnarray}
and
\begin{equation}
\delta\phi^{\mu,k}= ie^k_{\phi}\rho^{\alpha}\partial_{\alpha}X^{\mu}\epsilon.
\end{equation}
with $\epsilon$ is the infinetesimal constant anticommutating spinor. The commutator of
two transformations is the translation
\begin{equation}
[ \delta_1 ,\delta_2 ] X^{\mu} = a^{\alpha}\partial_{\alpha}X^{\mu},
[ \delta_1 ,\delta_2 ] \psi^{\mu j} = a^{\alpha}\partial_{\alpha}\psi^{\mu j}\nonumber\\
\end{equation}
and\\
\begin{equation}
 [ \delta_1 ,\delta_2 ] \phi^{\mu k} = a^{\alpha}\partial_{\alpha}\phi^{\mu k}\label{2}
\end{equation}
where $a^{\alpha}=2i\bar{\epsilon}_1\rho^{\alpha}\epsilon_2$
Equation(\ref{2}) is true only and only if
\begin{equation}
\psi_j^{\mu} = e_{\psi j}\Psi^{\mu}\nonumber\\
\end{equation}
and
\begin{equation}
\phi_k^{\mu} = e_{\phi k}\Psi^{\mu}
\end{equation}
The Majorana fermion
\begin{equation}
\Psi^{\mu} = e_{\psi}^j\psi^{\mu}_j - e_{\phi}^k\phi^{\mu}_k
\end{equation}
is such that
\begin{eqnarray}
\delta X^{\mu}=\bar{\epsilon}\Psi^{\mu}\nonumber\\
\delta \Psi^{\mu}=-i\epsilon\rho^{\alpha}\partial_{\alpha}X^{\mu}
\end{eqnarray}
and
\begin{equation}
[\delta_1 ,\delta_2]\Psi^{\mu } = a^{\alpha}\partial_{\alpha}X^{\mu }
\end{equation}
and is the right candidate to be the supersymmetric partner of the bosonic 
coordinates $X^{\mu}$ and also is the desired sum of $\psi^{\mu}$. The action in (1)
is only world sheet supersymmetric, but not local 2d supersymmetric.

Besides the above  pair, with the introduction of another pair, the zweibein
$e_{\alpha}(\sigma ,\tau)$ and the two dimensional spinor-cum-world sheet vector 
$\chi_{\alpha}$, $\delta\chi_{\alpha}=\nabla_{\alpha}\epsilon$ the local 2d supersymmetric 
invariant action is ~\cite{gr}
 
\begin{equation}
S= -\frac{1}{2\pi}\int d^2\sigma ~~e~~\left [ h^{\alpha\beta}\partial_{\alpha}X^{\mu }
\partial_{\beta}X_{\mu } -i\bar \Psi^{\mu}\rho^{\alpha}\partial_{\alpha}
\Psi_{\mu}+ 2\bar{\chi}_{\alpha}\rho^{\beta}\rho^{\alpha}\Psi^{\mu}
\partial_{\beta}X_{\mu}+\frac{1}{2}
\bar{\Psi }^{\mu}\Psi_{\mu}\bar{\chi}_{\beta} \rho^{\beta}\rho^{\alpha}\chi_{\alpha}\right ].
\label{s1}
\end{equation}
This is necessary to derive the vanishing of the energy momentum tensor and Noether current
from a gauge principle~\cite{gr,br}.
Varying the field and the zweibein, we derive the vanishing of the supercurrent 
$J_{\alpha}$ and the energy momentum tensor $T_{\alpha\beta}$ on the world sheet,
\begin{equation}
J_{\alpha}= \frac{\pi}{2e}\frac{\delta S}{\delta\bar{\chi}^{\alpha}}=\rho^{\beta}
\rho_{\alpha}\bar{\Psi}^{\mu}
\partial_{\beta}X_{\mu} -\frac{1}{4}
\bar{\Psi }^{\mu}\Psi_{\mu}\rho_{\alpha}\rho^{\beta}\chi_{\beta} =0
\end{equation}
The gravitino field $\chi_{\alpha}$ decouples and in the gauge $\chi_{\alpha}$=0,
we have
\begin{equation}
J_{\alpha}=\rho^{\beta}\rho_{\alpha}\bar{\Psi}^{\mu}\partial_{\beta}X_{\mu}=0.
\end{equation}
Variation of action with respect to `zweibein' ~~$e^a_{\alpha}$ ~~gives 
~~$T_{\beta}^a=-\frac{\pi}{2e}\frac{\delta S}{\delta e^{\beta}_a}$ = 0 and implies~~
$T_{\alpha\beta}=e_{\alpha a}T_{\beta}^a$. ~~~The variation with respect to the factor 
~~`e'~~ gives the traceless part. Since ~~$h^{\alpha\beta}~~
=~~e^{\alpha}_a e^{\beta}_b \eta^{ab}$, the bosonic term is obtained by finding the 
variation $\frac{\delta S}{\delta h^{\alpha\beta}}$.~~Due to spin connection~~
$\nabla_{\alpha}$~~ of the fermionic part is simply ~~$\partial_{\alpha}$~~ and
$\rho^{\alpha}\partial_{\alpha}$~~ should be written as ~~$\rho^a e_a^{\beta}
\partial_{\beta}$~~ so that the variation gives~~~$ S^F~~\sim~~ \frac{ie}{2}
\bar{\Psi}^{\mu}\rho^a~e^{\alpha}_a\partial_{\alpha}\Psi_{\mu}$~~ 
leading to ~~$\frac{\delta S^F}{\delta e^a_{\alpha}}~~\sim~~\frac{ie}{2}
\bar{\Psi}^{\mu}\rho^a\partial_{\alpha}\Psi_{\mu}$~~. The result is
\begin{equation}
T_{\alpha\beta}= - \frac{\pi}{2e}e_{\alpha a} T^a_{\beta} =
\partial_{\alpha}X^{\mu }
\partial_{\beta}X_{\mu }- \frac{i}{2}\bar{\Psi}^{\mu}\rho_{(\alpha}\partial_{\beta )}
\Psi_{\mu}-(trace)=0
\end{equation}
In a basis $\tilde {\psi}=(\psi^+,\psi^-)$ and $\tilde {\phi}=(\phi^+,\phi^-)$
the vanishing of the light cone components are
\begin{equation}
J_{\pm}=\partial_{\pm}X_{\mu}\Psi^{\mu}_{\pm}\label{J}
\end{equation}
and
\begin{equation}
T_{\pm\pm}=
\partial_{\pm}X^{\mu}\partial_{\pm}X_{\mu}+\frac{i}{ 2}{\psi}^{\mu j}_{\pm}\partial_{\pm}
 \psi_{\pm\mu,j }- \frac{i}{2}{\phi}_{\pm}^{\mu k}\partial_{\pm}\phi_{\pm\mu,k}
\end{equation}
where\\
$\partial_{\pm}=\frac{1}{2}(\partial_{\tau} \pm\partial_{\sigma})$. The component constraints
are the following equations (\ref{J})
\begin{eqnarray}
\partial_{\pm}X_{\mu}\psi_{\pm}^{\mu,j} = \partial_{\pm}X_{\mu}
e^j_{\psi}\Psi^{\mu}_{\pm }=0,~~~~~~~~j=1,2...6.\\ 
\partial_{\pm}X_{\mu}\phi_{\pm}^{\mu k} = \partial_{\pm}X_{\mu}e^k_{\phi}\Psi^{\mu}
_{\pm }=0,~~~~~~~~k=1,2...5.
\end{eqnarray}
These eleven constraints eliminate the eleven Lorentz metric ghosts 
from the physical spectrum. However they follow from one current constraint of equation
~(\ref{J}). Besides the conformal ghosts(b,c), it appears, we shall need eleven pairs of the
super conformal $(\beta^j, \gamma^j)$  and $(\beta^k, \gamma^k)$ ghosts to write down
BRST charge. These eleven pairs behave like one pair $(\beta, \gamma)$, like the current 
generator. One can easily construct `null' physical states as done in reference~\cite{gr}. 

For the two groups of Fermions the general constraints 
are the two equations
\begin{equation}
\partial_{\pm}X_{\mu}e^k_{\phi}\phi^{\mu}_{\pm k} = \partial_{\pm}X_{\mu}
e^j_{\psi}\psi^{\mu}_{\pm j}=0\label{5}
\end{equation}

However action given in equation(\ref{s1}) is not invariant under four
dimensional local supersymmetry. To obtain the simplest Green Schwarz action for N=1
local supersymmetry, we note that there are four component Dirac 
spinor representations of SO(3,1) which we denote by $\theta_{j\delta}$ and $\theta_{k\delta}$
respectively with $\delta$=1,2,3,4. A genuine space time fermion can be constructed rather 
than space time vector as
\begin{equation}
\theta_{\delta}=\sum^6_{j=1}e^j_{\psi}\theta_{j\delta} -
\sum^5_{k=1}e^k_{\phi}\theta_{k\delta}
\end{equation}
The G.S. action is~\cite{gr} for N=1
\begin{equation}
S=\frac{1}{2\pi}\int d^2\sigma \left ( \sqrt{g}g^{\alpha\beta}\Pi_{\alpha}\Pi_{\beta}
+2i\epsilon^{\alpha\beta}\partial_{\alpha}X^{\mu}\bar{\theta}\Gamma_{\mu}
\partial_{\beta}\theta\right )
\end{equation}
where $\Gamma_{\mu}$ are the Dirac gamma matrices and
\[ \Pi^{\mu}_{\alpha}=\partial_{\alpha}X^{\mu}- i\bar{\theta}\Gamma^{\mu}\partial_
{\alpha}\theta\].
The action is space time supersymmetric in four dimensions~\cite{gr}. We have traded the
vector four indices $\mu$ for the four spinor components $\delta$ to form the genuine
space time Dirac Spinors..
The major defect is that the naive covariant quantisation does not work. To proceed with 
covariant formulation we have to implement NS-R~\cite{ne} scheme with G.S.O.~\cite{gl}
projection which is simple, elegant and equivalent.
The non vanishing equal time commutator and anti commutators follow from the action 
in equation(1).
\begin{equation}
\left [\partial_{\pm}X^{\mu}(\sigma ,\tau), \partial_{\pm}X^{\nu}(\sigma' ,\tau)
\right ] = \pm\frac{\pi}{2} \eta^{\mu\nu}\delta'(\sigma-\sigma')
\end{equation}
and
\begin{equation}
\{ \psi_A^{\mu}(\sigma) , \psi_B^{\nu}(\sigma')\} =
\pi \eta^{\mu\nu}\delta'(\sigma-\sigma')\delta_{AB}
\end{equation}
Even though
\begin{equation}
\{ \phi_A^{\mu}(\sigma) , \phi_B^{\nu}(\sigma')\} =-
\pi \eta^{\mu\nu}\delta'(\sigma-\sigma')\delta_{AB},
\end{equation}
A,B refer to the component indices 1,2 or the helicities `+,$-$'.
There are no ghost quanta other than $\mu =\nu =0$ due to negative phase of 
creation operator. We immediately check that
\begin{equation}
[ T_{++}(\sigma),  T_{++}(\sigma') ] =i\pi\delta'(\sigma-\sigma') [T_{++}(\sigma)+
 T_{++}(\sigma') ]
\end{equation}
\begin{equation}
[T_{++}(\sigma),  J_{+}(\sigma')] =
i\pi\delta'(\sigma-\sigma') [ J_{+}(\sigma)+\frac{1}{2} J_{+}(\sigma') ]
\end{equation}
and
\begin{equation}
[J_{+}(\sigma),  J_{+}(\sigma')] =\pi\delta'(\sigma-\sigma')T_{++}(\sigma)
\end{equation}
Similarly for $-~-$ , $-$ components. The superalgebra closes.

Let $\alpha_m^{\mu}$ denote the quanta of coordinates satisfying
\begin{equation}
[\alpha_m^{\mu}, \alpha_n^{\nu}]=m\delta_{m,-n}\eta^{\mu\nu}
\end{equation}
With NS conditions, let $b^{\mu}_r $ and $b^{'\mu}_r$ be the quanta of $\psi$ and 
$\phi$ fermions with half integral r where as with Ramond boundary conditions 
they are $d^{\mu}_m$ and $d^{'\mu}_m$ with m integers. The non vanishing 
anticommutator relations are
\begin{equation}
\{ b^{\mu ,j}_r , b^{\nu,j'}_s\}=\eta^{\mu\nu}\delta^{j,j'}\delta_{r,-s} ;~~~~~~~~~~
~~~~~\{ b'^{\mu ,k}_r , b'^{\mu,k'}_s\}=\eta^{\mu\nu}\delta^{k,k'}\delta_{r,-s}
 ~~~:~~~~~~~~NS
\end{equation}
\begin{equation}
\{ d^{\mu ,j}_m , d^{\nu,j'}_n\}=-\eta^{\mu\nu}\delta^{j,j'}\delta_{m,-n} ;~~~~~~~~~
~~~~~\{ d'^{\mu ,k}_m, d'^{\nu,k'}_n\}=-\eta^{\mu\nu}\delta^{k,k'}\delta_{m,-n} ~~~~:~~~~~~R
\end{equation}
with
\begin{equation}
b^{\prime\mu}_{-r,j}=-b^{\prime\dag\mu}_{r,j}\;\;\;\;\;\;\; and\;\;\;\;\;\;\;\;\;\;
d^{\prime\mu}_{-m,j}=-d^{\prime\dag\mu}_{m,j}
\end{equation}
It will be useful to note ~~$ b_r^{\mu,j} = e^j_{\psi} B_r^{\mu},~~~
b_r^{\prime\mu,k} = e^k_{\phi} B_r^{\mu}, ~~~~B_r^{\mu} = 
e^j_{\psi}b_{r,j}^{\mu} - e^k_{\phi} b_{r,k}^{\prime\mu}$.~~Similarly for the d, d';\\ 
$D_m^{\mu} = e^j_{\psi}d_{m,j}^{\mu} - e^k_{\phi}b_{m,k}^{\prime\mu}$.

Virasoro generators~\cite{vr}
\begin{eqnarray}
L_m &=&\frac{1}{\pi}\int_{-\pi}^{\pi}d\sigma e^{im\sigma}T_{++}\\
G_r &=&\frac{\sqrt{2}}{\pi}\int_{-\pi}^{\pi}d\sigma e^{ir\sigma}J_{+} ~~~~~~~~:~~~~~~~~~~~~~~ NS\\
F_n& =&\frac{\sqrt{2}}{\pi}\int_{-\pi}^{\pi}d\sigma e^{in\sigma}J_{+}~~~~~~~~ :~~~~~~~~~~~~~~~ R\\
\left [L_m , L_n \right ]&=&(m-n)L_{m+n} +A(m)\delta_{m,-n}\\
\left [L_m , G_r \right ]&=&(\frac{1}{2}m-r)G_{m+r}~~~~~~~~~~~~~:~~~~~~~~~~~~~~~NS \\
\{ G_r , G_s \}&=&2L_{s+r} +B(r)\delta_{r,-s}\\
\left [L_m , F_n \right ]&=&(\frac{1}{2}m-n)F_{m+n}~~~~~~~~~~~~~:~~~~~~~~~~~~~~~R\\
\{F_m , F_n \}&=&2L_{m+n} +B_F(m)\delta_{m,-n}
\end{eqnarray}
Here A(m), B(r) and $B_F(m)$ are normal ordering constants, since the generators are
the sum of the products of normal ordered quantum operators `: :'. A single dot 
implies sum over all qualifying indices .

\begin{equation}
NS:~~~~~~~~~~L_m = \frac{1}{2}\sum^{\infty}_{-\infty}:\alpha_{-n}\alpha_{m+n}: 
+ \frac{1}{2} \sum^{\infty}_{-\infty}(r+\frac{1}{2}m): (b_{-r} 
\cdot b_{m+r} - b_{-r}' \cdot b_{m+r}'): 
\end{equation}
\begin{equation}
G_r= \sum^{\infty}_{-\infty}\alpha_{-n}\cdot B_{n+r}
\end{equation}
\begin{equation}
R:~~~~~~~~~`L_m = \frac{1}{2}\sum^{\infty}_{-\infty}:\alpha_{-n}\alpha_{m+n}: 
+ \frac{1}{2} \sum^{\infty}_{-\infty}(n+\frac{1}{2}m): (d_{-n} \cdot d_{m+n} - d_{-n}' 
\cdot d_{m+n}'):
\end{equation}

\begin{equation}
F_m= \sum^{\infty}_{-\infty}\alpha_{-n}\cdot D_{n+m}
\end{equation}
Both in NS and R sectors, there are the light cone ghosts corresponding to the 
fermionic components $\mu =0$ and 3 which are present in $L_m$.
By standard technique, we calculate that
\begin{equation}
A(m)=\frac{26}{12}(m^3-m)=\frac{C}{12}(m^3-m).
\end{equation}
C=26 is the central charge. Using Jacobi Identity
\begin{equation}
B(r)=\frac{1}{2r}A(r)=\frac{26}{3}(r^2-\frac{1}{4})=\frac{C}{3}(r^2-\frac{1}{4}).
\end{equation}
The central charge C can also be deduced from the v.e.v. of the product of two energy
momentum tensors e.g.
\begin{equation}
< T_+(z)T_+(\omega)>\sim \frac{C}{2}(z-\omega)^{-4}+.........
\end{equation}
with
\begin{equation}
C=\eta^{\mu}_{\mu}+\frac{1}{2}\eta^{\mu}_{\mu}\delta^j_j +
\frac{1}{2}\eta^{\mu}_{\mu}\delta^k_k=26.
\end{equation} 
The normal order anomally in Ramond sector must note that $F_o$ has no ambiguity. 
So $F_o^2 =L_o$. Using the algebraic relation for F's and the Jacobi Identity
\begin{equation}
\{ F_r , F_{-r}\} = \frac{2}{r}\{ [L_r ,F_o], F_{-r}\} =L_o + \frac{4}{r} A(r)
\end{equation} 
leads to
\begin{equation}
 B_R(m)=\frac{4}{m} A(m)=\frac{C}{3}(m^2 - 1);~~~~~~~~~~ m\neq 0
\end{equation}
The anomalies are have to be cancelled out and one needs the Faddeev Popov
ghosts $b_{\pm\pm}$, $c^{\pm}$ fields given by the action
\begin{equation}
S_{FP}= \frac{1}{\pi}\int d^2\sigma (c^+\partial_-b_{++}+c^-\partial_+b_{--})
\end{equation}
The $c^{\pm},  b_{\pm\pm}$ field quanta $b_n, c_n$ satisfy the anticommutator relation
$\{c_m, b_n\}= \delta_{m,-n} $ and $\{c_m, c_n\}= \{b_n, b_m\}=0 $
The Virasoro generators for the ghost become
\begin{equation}
L_m^{gh}=\sum^{\infty}_{n=-\infty}(m-n)b_{m+n}c_{-n}-a\delta_{m,0}
\end{equation}
where a is the normal ordering constant. If $G_r^{gh}$ and $F_r^{gh}$ are 
the ghost current generators of conformal dimension $\frac{3}{2}$, the super 
Virasoro algebras for the ghosts for the two sectors follow. The anomaly 
term is deduced to be
\begin{equation}
A^{gh}(m)=\frac{1}{6}(m-13m^3) +2am=-\frac{26}{12}(m^3-m) +2(a-1)m
\end{equation}
To be anomaly free, $A^{gh}(m)$ +A(m)=0. This gives the value of a=1 for the normal 
ordering constant and -26 for the central charge of the ghost. Due to Jacobi 
Identity, the anomaly terms for the ghosts for both the sectors cancels as well.
Superconformal ghosts which contributes central charge 11 in a normal superstring
are not necessary. To explain this we observe that the light cone gauge is
ghost free.Dropping the helicity suffixes, the lightcone vectors,
$\psi^{\pm}_j=\frac{1}{\sqrt{2}}(\psi_j^o \pm \psi^{3}_j )$ and
$\phi^{\pm}_j=\frac{1}{\sqrt{2}}(\phi_j^o \pm \phi^{3}_j) $ have the 
anticommutators with negative sign and are the ghost modes. The total ghost 
energy-momentum tensor comes from these (0,3) coordinates.
\begin{equation}
T^{gh}(z)= \frac{i}{2}( \psi^{0j}\partial_z\psi_{0j}  +
\psi^{3j}\partial_z\psi_{3j})
-\frac{i}{2}( \phi^{0k}\partial_z\phi_{0k}  +
\psi^{3k}\partial_z\psi_{3k} )
\end{equation}
But the vacuum correlation function is
\begin{equation}
<T^{gh}(z)T^{gh}(\omega)> = \frac{11}{2} (z-\omega)^{-4}+ \cdot\cdot
\end{equation}
Thus the contribution of these ghosts to the central charge 26 is 11 like 
the superconformal ghosts. Without these ghosts the central charge is 15 as is the
case for normal ten dimensional superstring.
With no anomalies and ghosts, the physical states can be constructed as
\begin{equation}
Bosonic~~~~~~~~~NS:~~~~~~ (L_o-1)|\Psi> =0,~~~~~~L_m|\Psi>=0,~~~~~G_r|\Psi>=0;~~~m,r>0
\label{56} 
\end{equation}
\begin{equation}
Fermionic~~~~~~~R:~~~~~~ (L_o-1)|\Phi>_{\alpha} =0,~~~~~~L_m|\Phi>_{\alpha}=0,
~~~~~F_m|\Phi>_{\alpha}=0;~~~m>0\label{57}
\end{equation}
Since $L_o = F_o^2$, we also have ($F_o-1)|\phi_+>_{\alpha}=0$ and $(F_o+1)|\phi_->_{\alpha}$=0.

We proceed to prove that these physical state conditions eliminate all the ghosts. We
begin with Gupta Bleuler postulate  $\alpha^o_{-1}|\Phi>=0$. From the physical state
condition for $L_m$, ~~$[L_{m+1}, \alpha^o_{-1}]|\Phi>=\alpha^o_m|\Phi>$=0~~. 
The longitudinal modes of the coordinates drop out. For the bosonic NS sector,
the physical state condition for $B_r$ is used.~~$[ B_{r+1}, \alpha^o_{-1}]|\Phi>=
B_r^o|\Phi>=0$~~. By definition, ~~$b_{r}^{j,o}|\Phi>=e_{\psi}^jB_r^o|\Phi>=0$~~.
Similarly,~~$b_{r}^{'k,o}|\Phi>=e_{\phi}^k B_r^o|\Phi>=0$.~~~ In the fermionic R sector, the physical state condition for ~~$F_m$,~~ leads to
~~$[F_{m+1}, \alpha^o_{-1}]|\Phi>=D^o_m|\Phi>$=0.~~As before, by definition, 
~~~$d_m^{ko}|\Phi> = d_m^{'ko}|\Phi> = 0$.~~Thus the longitudinal modes of all the 
quanta are absent in the Fock space. Using the usual pair of $(\beta,\gamma)$ 
commuting ghosts, the nilpotent BRST charge operator
\begin{equation}
Q_{BRST}=\sum L_{-m}c_m -\frac{1}{2}\sum (m-n):c_{-m}c_{-n}b_{m+n}: - c_o + Q'
\end{equation}
where ~~$L_{-m}$~~ is given by equation (42) and
\begin{equation}
 Q'=\sum G_{-r}\gamma_r -\sum\gamma_{-r}\gamma_{-s}b_{r+s}~~~~~~~~~~~~for~~  NS;
\end{equation}
~~$L_{-m}$~~ is given by the equation (44) and
\begin{equation}
 Q'=\sum F_{-n}\gamma_{n} -\sum\gamma_{-n}\gamma_{-m}b_{n+m}.~~~~~~~~~~ for~~~ R
\end{equation}
Q' takes care of the constraints due to the vanishing of the current generator operators
acting on the physical states ~(\ref{56}) and~(\ref{57}). In constructing the nilpotent
charge we strictly followed the procedure for construction from operators satisfying
Lie and graded Lie algebra [6].It is to be noted that $Q_{BRST} - Q'$ and $Q_{BRST}$ are 
separately nilpotent. 

Actually there are eleven subsidiary physical state conditions 
like  ~(\ref{56}) and~(\ref{57}) having the 
property following the fourier transformation and definition
\begin{equation}
G_r^j|\Phi> = e_{\psi}^j G_r |\Phi> = 0; ~~~~~~~~G_r^k|\Phi> = e_{\phi}^k G_r |\Phi> = 0 ~~~~~~~~NS
\end{equation}
\begin{equation}
F_m^j|\Phi> = e_{\psi}^j F_m |\Phi> = 0; ~~~~~~~ F_m^k|\Phi> = e_{\phi}^k F_m |\Phi> = 0 ~~~~~~~~R,
\end{equation} 
with $ G_r =  e_{\psi}^j G_{r,j} -  e_{\phi}^k G_{r,k}~~ and~~  
F_m =  e_{\psi}^j F_{m,j} -  e_{\phi}^k F_{m,k}$. We can follow up by writing $(\gamma^j_r,
\beta^j_r)$ and  $(\gamma^k_r, \beta^k_r)$ with similar definitions. But, in the NS sector,
 for instance, $ G^j_{-r}\gamma_{r,j} - G^k_{-r}\gamma_{r,k}\equiv G_{-r}\gamma_r$ where
$\gamma_r = e^j_{\psi}\gamma_{r,j} -  e^k_{\phi}\gamma_{r,k}$. So that the expressions for Q
and $Q'$  contain the conformal ghosts (b,c) and the pair of ($\beta ,\gamma$) super
conformal ghosts unitedly allowing for a BRST charge $ Q^2_{BRST}$= 0. The theory is ghost
free and unitary.

$L_o$, being the Hamiltonian, we obtain the mass spectrum of the model
\begin{equation}
NS:~~~~~~~~~~~~~~~~~~~~~~~~\alpha'M^2=-1,-\frac{1}{2},0,\frac{1}{2},1,\frac{3}{2},.....
\end{equation}
and 
\begin{equation}
R:~~~~~~~~~~~~~~~~~~~~~~~~~\alpha'M^2=-1,0,1,2,3,...........
\end{equation}

The self energy of the scalar tachyonic vacuum is cancelled by the contribution 
of the fermionic loop of the Ramond sector because the self energy of the fermion 
loop is $-<0|(F_o-1)^{-1}(F_o+1)^{-1}|0>= -<0|(L_o^{NS}-1)^{-1}|0>$.
 The half integral values are 
eliminated by the clever G.S.O projection of the states.
\begin{equation}
G=\frac{1}{2}(1+(-1)^{F+F'})
\end{equation}
where the F and $F'$ are the fermionic numbers of b and $b^{\prime}$ quanta. Thus both 
the fermions and bosons lie on linear Regge trajectories $\alpha'M^2=$ -1,0,1,2,3........ 
The G.S.O. projections is necessary and essential to prove the modular
invarianc of the partition function of this string. 

In covariant formulation, 
the effective number of the physical modes is the total number of modes minus the
number of invariant constraints. Here there are forty four fermions and four
 constraints given by equation (\ref{5}). So there are forty space time fermions in
the assembly. Let us cast them in SO(6) and SO(5) invariant way.
Let $b^k_{r,\alpha}$ be the annihilation quanta. $\alpha$ runs from 1 to 8 and 
k from 1 to 5. In the NS sector the Hamiltonian $H^{NS}=\sum_{k=1}^5H_k^{NS}$ where
$H^{NS}_k=\sum_r b^{k\dagger}_{r,\alpha}b^{k}_{r,\alpha}-\frac{8}{48}$. But
$[ H_k^{NS}, H_{k'}^{NS}]=0$. The partition function of the forty member 
assembly will be the partition function of eight fermions  raised to the power of 
five.
The path integral function for the eight fermions is given by Seiberg and Witten
~\cite{se}. In their notation 
\begin{eqnarray}
A_k ((-~-),\tau)= (\frac{\Theta_3 (\tau)}{\eta(\tau)})^4\\
A_k((+~-),\tau) = A_k((-~-),\frac{\tau }{(1+\tau)})=- 
(\frac{\Theta_2(\tau)} {\eta(\tau)})^4\\
A_k((-~+),\tau)=A_k((+~-),\frac{-1}{\tau})=- 
(\frac{\Theta_4(\tau)}{\eta(\tau)})^4\\
A_k((+~+),\tau)=0
\end{eqnarray}
 The sum of all spin structure is 
\begin{equation}
A_k(\tau) =\frac{\Theta^4_3(\tau)- \Theta^4_2(\tau)-\Theta^4_4(\tau)}
{\eta^4(\tau)}
\end{equation}

The total partition function $Z=|A_k(\tau)|^{10}$ is not only modular invariant but 
vanishes due to the famous Jacobi relation which is the necessary condition for 
space time supersymmetry.

Thus we have been able to construct a novel modular invariant superstring which 
is free of anomalies and ghosts. Novel features of the model are that this is the only
superstring known and constructed whose central charge is 26 and  there is no need for
superconformal ghosts in covariant formulation except in the construction of BRST Charge.
These ghosts are usually contributing 11 to the central charge and are replaced by 11 
Lorentz metric ghost of vector fermions. This linkage needs further study. There are 
useful and harmless tachyons in NS and R sector, but the self energies of the tachyonic 
vacua cancel.This cancellation mechanism, perhaps, makes the vacua unique.
Since bosonic string theory is simpler and well understood in general, a derivation of the 
four dimensional space time supersymmetric action will positively help in studying the 
dynamics of the superstring theory in greater depth and insight.


\begin{thebibliography}{99}
\bibitem{Int} I. Antoniadis, C.Bachas and C Kounnas, Nucl.Phys.{\bf B289}(1987)87
\bibitem{ch} D Chang and A. Kumar, Phys. Rev. {\bf D38}(1988)1893; ibid {\bf D38}
(1988) 3739
\bibitem{ka} H. Kawai, D.C. Lewellen and C-H. H. Tye Nucl. Phys. {\bf B288}(1987) 1
\bibitem{cas} A. Casher, F. Englert, H. Nicolai and A. Taormina, 
Phys. Lett. {\bf B162}(1985)121 
\bibitem{en} F. Englert, L. Houart and A Taormina, JHEP, 0108.013(2001)1
\bibitem{gr} For notation, reference and details, here and elsewhere, see the 
text books M.B. Green, J. H. Schwarz and E. Witten,\\ 
{\it Superstring Theory} Vol-I, Cambridge University Press, Cambridge, England(1987) and 
M. Kaku, {\it Introduction to Superstring and M-Theory}, Springer Verlag, New York(1988)
\bibitem{ne} A. Neveu and J.H. Schwarz, Nucl. Phys. {\bf B31}(1971)86; P Ramond,
Phys. Rev. {\bf D3}(1971) 2415
\bibitem{gl} F. Gliozzi, J. Scherk and D. Olive, Phys. Lett.{\bf 65B}(1976) 282
\bibitem{vr} M. A. Virasoro, Phys. Rev.{\bf D1}(1970) 2933
\bibitem{se} N. Seiberg and E. Witten, Nucl. Phys. {\bf B276}(1986)27
\bibitem{br} L. Brink, P. Di Vecchia and P. Howe, Phys. Lett. {\bf 65B}(1976)471;
S. Deser and B. Zumino, Phys. Lett. {\bf 65B}(1976)369.
\end{thebibliography}
\end{document}